\documentclass[useAMS,usenatbib]{mn2e}

\usepackage[psamsfonts]{amssymb}
\usepackage[dvips]{graphicx}
\usepackage{graphicx}
\usepackage{amsmath,alltt}                                                                                                                          
\usepackage{multirow}
\usepackage{rotating}
\usepackage{lscape}
\usepackage{tablefootnote}
\usepackage{hyperref}
\title[Energy Equipartition and the Stellar Mass Function]{On the Link Between Energy Equipartition and Radial Variation in the Stellar Mass Function of Star Clusters}
\author[Webb $\&$ Vesperini]{Jeremy J. Webb \& Enrico Vesperini
\thanks{E-mail: jerjwebb@iu.edu (JW), evesperi@indiana.edu (EV)} \\
Department of Astronomy, Indiana University, Swain West, 727 E. 3rd Street, IN 47405 Bloomington, USA}

\begin{document}

\pagerange{\pageref{firstpage}--\pageref{lastpage}} \pubyear{2015}

\maketitle

\label{firstpage}

\begin{abstract}

We make use of $N$-body simulations to determine the relationship between two observable parameters that are used to quantify mass segregation and energy equipartition in star clusters. Mass segregation can be quantified by measuring how the slope of a cluster's stellar mass function $\alpha$ changes with clustercentric distance r, and then calculating $\delta_\alpha = \frac{d \alpha(r)}{d ln(r/r_m)}$ where $r_m$ is the cluster's half-mass radius. The degree of energy equipartition in a cluster is quantified by $\eta$, which is a measure of how stellar velocity dispersion $\sigma$ depends on stellar mass m via $\sigma(m) \propto m^{-\eta}$. Through a suite of $N$-body star cluster simulations with a range of initial sizes, binary fractions, orbits, black hole retention fractions, and initial mass functions, we present the co-evolution of $\delta_\alpha$ and $\eta$. We find that measurements of the global $\eta$ are strongly affected by the radial dependence of $\sigma$ and mean stellar mass and the relationship between $\eta$ and $\delta_\alpha$ depends mainly on the cluster's initial conditions and the tidal field. Within $r_m$, where these effects are minimized, we find that $\eta$ and $\delta_\alpha$ initially share a linear relationship. However, once the degree of mass segregation increases such that the radial dependence of $\sigma$ and mean stellar mass become a factor within $r_m$, or the cluster undergoes core collapse, the relationship breaks down. We propose a method for determining $\eta$ within $r_m$ from an observational measurement of $\delta_\alpha$. In cases where $\eta$ and $\delta_\alpha$ can be measured independently, this new method offers a way of measuring the cluster's dynamical state.

\end{abstract}

\begin{keywords}
methods: statistical Ð stars: statistics Ð globular clusters: general
\end{keywords}

\section{Introduction} \label{intro}

Two-body gravitational interactions between stars within star clusters is one of the main dynamical processes governing their long-term evolution. Over time, two-body interactions result in the stellar population evolving towards a state of energy equipartition on a timescale commonly referred to as a cluster's relaxation time \citep{spitzer87}. A natural by-product of two-body relaxation is that the cluster also becomes mass segregated. Mass segregation occurs because through two-body interactions, massive stars lose kinetic energy to less massive stars and fall inwards towards the centre of the cluster. Conversely, lower mass stars that have had their kinetic energy increased migrate outwards. Therefore the effects of two-body relaxation are expected to imprint a mass dependence in both the kinematical properties and the radial distribution of stars within a star cluster. The strength of this mass dependence evolves with the dynamical age of a cluster, and should be larger for clusters in the more advanced stages of their dynamical evolution.

In a star cluster which has reached a state of complete energy equipartition, stars at a given distance from the cluster centre will have a mass dependent velocity dispersion scaling as  $\sigma(m) \propto m^{-\eta}$, with $\eta=0.5$. Some simplified two-component systems have been shown to reach $\eta=0.5$, however the conditions must be idealized such that the cluster does not under-go core collapse with heavier stars becoming self-gravitating \citep{spitzer69, spitzer71, lightman78, inagaki84}. On the other hand, a number of studies have shown that for systems containing stars with a spectrum of masses following realistic stellar initial mass functions (IMFs) complete energy equipartition will never be reached \citep{vishniac78, inagaki85, watters00, fregeau02, baumgardt03, khalisi07, parker16, spera16}. In a recent study based on direct N-body simulations of multi-mass star clusters, \citet{trenti13} explored a wide range of initial conditions and IMFs and found that in no case complete energy equipartition was reached. The authors also find that the inner region of a cluster can get closer to energy equipartition than the outer region ($\eta_{max} \sim 0.15$), but after many relaxation times stars in different radial bins converge to a similar state of "partial equipartition" ($\eta \sim 0.08$). A similar conclusion was reached by \citet{bianchini16} using a set of Monte Carlo cluster simulations. While measurements of $\sigma(m)$ are becoming possible with Hubble Space Telescope proper motion studies and data from GAIA \citep[e.g.][]{wright16}, measurements of the dependence of kinematical properties on mass are still challenging \citep[e.g.][]{baldwin16}.

Since mass segregation is a direct consequence of a cluster's evolution towards energy equipartition, the degree of mass segregation within a cluster should be connected with the level of energy equipartition \citep{trenti13}. Observational studies of globular clusters have recently been able to study how the radial distribution of stars depends on stellar mass by looking at how the slope of cluster's stellar mass function ($\alpha$) changes with projected clustercentric distance ($R$) \citep{beccari10, frank12, frank14, dalessandro15, beccari15, zhang15}. With the exception of NGC 6101 \citep{dalessandro15}, the behaviour of $\alpha(R)$ in each cluster is such that the mass function is flatter (top-heavy) in the inner regions and steeper (bottom-heavy) in the outer regions of the cluster indicating the clusters are at least partially mass segregated. 

In \citet{webb16} we have quantified the radial variation in a cluster's mass function as $\delta_\alpha$, the slope of a plot of $\alpha$ versus ln$(\frac{r}{r_m})$ where $r$ and $r_m$ are the clustercentric distance and half-mass radius respectively (presented in three-dimensions and in projection). We found that $\delta_\alpha$ decreases from zero (no segregation) with time as a cluster segregates, with the rate at which $\delta_\alpha$ decreases primarily depending on a cluster's initial half-mass relaxation time $t_{rh}$ and tidal filling factor $\frac{r_m}{r_t}$ (where $r_t$ is the cluster's tidal radius). We also find that $\delta_\alpha$ reaches a similar minimum value for all our model clusters such that clusters never become fully mass segregated. We suggest that the existence of a minimum $\delta_\alpha$ must be linked to the overwhelming evidence that clusters can only reach a minimum value of $\eta$ that is greater than -0.5.

The purpose of this study is to use $N$-body simulations of star clusters with a range of initial conditions to study the co-evolution of a cluster's $\delta_\alpha$ and $\eta$ (both three dimensionally and in projection). Linking observational measurements of the degree of mass segregation in a cluster to its state of energy equipartition would provide a powerful tool for indirectly measuring energy equipartition in star clusters. In Section \ref{s_nbody} we summarize the initial conditions of the $N$-body models initially studied in \citet{webb16}. In Section \ref{s_evol} we determine how $\delta_\alpha$ and $\eta$ evolve for model clusters with different sizes, binary fractions, IMFs, black hole retention fractions, and orbits in a Milky Way-like tidal field. To make our results useful for observations, we also determine whether or not projection effects can alter the apparent evolution of either $\delta_\alpha$ or $\eta$. We discuss and summarize all of our findings in Section \ref{s_discussion}, where we introduce a method for inferring the value of $\eta$ based simply on a measurement of $\delta_\alpha$.

\section{N-body models} \label{s_nbody}

As originally discussed in \citet{webb16}, the evolution of each model star cluster in our suite of simulations was simulated using the direct $N$-body code NBODY6 \citep{aarseth03}. The initial positions and velocities of individual stars are generated assuming the cluster follows a Plummer density profile \citep{plummer11, aarseth74}. The Plummer profile is cut-off at a radius of 10 times the cluster's initial half-mass radius $r_{m,i}$, with model clusters having $r_{m,i}$ equalling either 1.1 pc or 6 pc. The IMF used to generate the initial distribution of stellar masses has the functional form:

\begin{equation}\label{eqn:mfunc}
\frac{dN}{dm}=m^\alpha
\end{equation}

\noindent where m represents stellar mass. For our main model clusters, the IMF is taken from \citet{kroupa93} (K93) where $\alpha$ equals -2.7 for $m > 1 M_\odot$, -2.2 for $0.5 \le m \le 1  M_\odot$, and -1.3 for $0.08 < m \le 0.5  M_\odot$. We restrict stellar masses to be between 0.1 and 50 $M_\odot$. The stellar evolution algorithm for individual stars is taken from \citet{hurley00} assuming all stars have the same metallicity (0.001). 

Using a K93 IMF, our base models taken from \citet{webb16} are $6\times10^4 M_\odot$ clusters with $r_{m,i}$ equalling 1.1 pc and 6 pc and circular orbits at 6 kpc in a Milky Way-like potential (RM1 and RM6 respectively). The Galactic potential consists of a $1.5 \times 10^{10} M_{\odot}$ point-mass bulge, a $5 \times 10^{10} M_{\odot}$ \citet{miyamoto75} disk (with $a=4.5\,$ kpc and $b=0.5\,$ kpc), and a logarithmic halo potential \citep{xue08}. The halo potential is scaled such that the circular velocity $8.5\,$ kpc from the Galactic centre is 220 km/s. Since each cluster starts with the same tidal radius $r_t$, the 1.1 pc cluster is initially tidally under filling ($\frac{r_m}{r_t} = 0.04$) and the 6 pc cluster is initially tidally filling ($\frac{r_m}{r_t}=0.2$). 

To study the effects of different IMFs on the evolution of $\delta_\alpha$ and $\eta$, model clusters with different IMFs from \citet{webb16} with $r_{m,i} = 6$ pc and circular orbits at 6 kpc in the Milky Way-like potential (identical to RM6) are considered. In addition to the K93 IMF, we consider a \citet{kroupa01} IMF (IMFK01), a Salpeter IMF with $\alpha$ =2.35 \citep{salpeter55} (IMF235), and a broken power law IMF with $\alpha$ equals -0.9 for $0.1 < m < 0.5 M_\odot$ and -2.3 for $0.5 \le m \le 50  M_\odot$ (IMFBPL). For non-K93 IMFs, the publicly available code McLuster \citep{kupper11} is used to generate stellar masses. 

To study a larger range of tidal filling factors, we make use of previous simulations from \citet{leigh13} and \citet{webb15} with circular orbits between 6 and 104 kpc, orbital eccentricities between 0 and 0.9, initial masses of $6 \times 10^4 M_\odot$, initial half mass radii of 1.1 pc and 6 pc, and primordial binary fractions of $4\%$. These models are named based on their eccentricity (E), circular orbit (R) or perigalactic (RP) distance, and initial $r_{m,i}$ (RM) such that, for example, a model cluster with an orbital eccentricity of 0.5, perigalactic distance of 6 kpc, and $r_{m,i}=1.1$ pc is named E05RP6RM1. To generate binary star masses, the sum of two stars drawn from the IMF is set equal to the total binary mass and the mass ratio is randomly taken from a uniform distribution. The stellar evolution algorithm for binary stars is taken from \citet{hurley02}, while the initial distribution of orbital eccentricities and periods are taken from  \citet{heggie75} and \citet{duquennoy91} respectively. When analyzing the models all binaries are treated as unresolved. 

In order to study the effects of black holes on the co-evolution of $\delta_\alpha$ and $\eta$, we ensure that a set fraction of all newly formed black holes are not given a velocity kick. For the purposes of this study we consider black hole retention fractions of $25\%$ (BH25) and $50\%$ (BH50). Table \ref{table:modparam} contains a summary of all the models discussed above.

\begin{table*}
  \caption{Model Input Parameters}
  \label{table:modparam}
  \begin{center}
    \begin{tabular}{lcccccccc}
      \hline\hline
      {Name} & {Mass} & {$r_{m,i}$} & {$R_c$} & {IMF} \\
      \hline

Star Clusters with Different Tidal Filling Factors $^1$ \\
\hline

RM1 & $6\times10^4 M_\odot$ & 1.1 pc & 6 kpc & K93 \\
RM6 & $6\times10^4 M_\odot$ & 6 pc & 6 kpc & K93 \\

\hline
Clusters with Different IMFs$^1$ \\
\hline

IMF235 & $6\times10^4 M_\odot$ &  6 pc & 6 kpc & $\alpha_0 = 2.35$ \\
IMFK01 & $6\times10^4 M_\odot$ & 6 pc & 6 kpc & K01 \\
IMFBPL & $6\times10^4 M_\odot$ &  6 pc & 6 kpc & $\alpha_0=-0.9$ for $0.1 < m < 0.5 M_\odot$ \\
{}&{}&{}&{}&$\alpha_0=-2.3$ for $0.5 \le m \le 50  M_\odot$. \\

\hline
Clusters with $4\%$ Initial Binaries and Different Orbits $^2$ \\
\hline

E0R6RM1 & $6\times10^4 M_\odot$ & 1.1 pc & 6 kpc & K93 \\
E05RP6RM1 & $6\times10^4 M_\odot$ & 1.1 pc & 6 kpc & K93 \\
E0R18RM1 & $6\times10^4 M_\odot$ & 1.1 pc & 18 kpc & K93 \\
E05RP6RM6 & $6\times10^4 M_\odot$ & 6 pc & 6 kpc &  K93 \\
E0R18RM6 & $6\times10^4 M_\odot$ & 6 pc & 18 kpc & K93 \\
E09RP6RM6 & $6\times10^4 M_\odot$ & 6 pc & 6 kpc & K93 \\
E0R104RM6 & $6\times10^4 M_\odot$ & 6 pc & 104 kpc & K93 \\

\hline
Star Clusters with Different Black Hole Retention Fractions$^1$ \\
\hline

BH25 & $6\times10^4 M_\odot$ & 6 pc & 6 kpc & K93 \\
BH50 & $6\times10^4 M_\odot$ & 6 pc & 6 kpc & K93 \\

      \hline\hline
      
\multicolumn{2}{l}{$^1$ \citet{webb16}} \\
\multicolumn{2}{l}{$^2$ \citet{webb13}, \citet{leigh13}, \citet{webb15}} \\
    \end{tabular}
  \end{center}
\end{table*}

\section{The Evolution of Energy Equipartition and Mass Segregation}\label{s_evol}
 
%From previous studies, we know that both $\eta$ and $\delta_\alpha$ will decrease with time as a cluster relaxes \citep{trenti13, bianchini16, webb16}. These studies also show that the time evolution of both parameters slows as a cluster expands and reaches near constant values at later times. If and how the evolution of $\eta$ correlates with that of $\delta_\alpha$ remains to be seen. We also note that \citet{trenti13} found that the evolution of $\eta$ is radially dependent, since high mass stars segregate faster than low mass stars and the core of a cluster relaxes quicker than the outer regions. The net result is that $\eta$ decreases rapidly in the inner regions of the cluster but slowly overall. Therefore we need to explore both the global coevolution of $\eta$ and $\delta_\alpha$ as well their evolution within different clustercentric radii r. And to compare the results of our study to observations of globular clusters, calculations of $\eta$ and $\delta_\alpha$ must also be done in projection. 
 
As described in \citet{webb16}, to measure $\delta_\alpha$ we first identify stars with masses within the desired mass range that the global mass mass function is going to be measured over. Commonly this mass range is either 0.1-0.5 $M_\odot$, 0.3-0.8 $M_\odot$, or 0.5-0.8 $M_\odot$ depending on the observational dataset. In this Section we will focus on the 0.1-0.5 $M_\odot$ mass range, however in Section \ref{s_discussion} all three mass ranges will be discussed. To measure the slope of the global mass function, $\alpha_G$, stars are first separated into 6 mass bins containing an equal number of stars. Keeping the bin width variable, such that each mass bin contains the same number of stars, allows for the cluster's mass function to determined with minimal numerical bias \citep{maiz05}. The value of m in each mass bin is set equal to the mean mass of stars within the mass bin. The slope of the line of best fit to $\frac{dN}{dm}$ versus $m$ is then  $\alpha_G$. 
 
To measure radial variation in the stellar mass function, we next separate the sub-population of stars within the given mass range into 10 radial bins. We again keep the width of each bin variable, such that each radial bin contains $10\%$ of the total sub-population. The distance of each bin from the centre of the cluster is set equal to the mean $r$ of all stars within the radial bin. After measuring $\alpha$ in each radial bin, $\delta_\alpha$ is then equal to the slope of the line of best fit to $\alpha$ versus ln$(\frac{r}{r_m})$. Repeating this process for each time step allows us to determine how $\delta_\alpha$ evolves with time for each model cluster (see \citet{webb16} for the evolution of $\delta_\alpha$ with respect to time and $\alpha_G$ for each model cluster in Table \ref{table:modparam}). 

To measure $\eta$ we first identify stars with masses between 0.1 and 1.8 $M_\odot$ (similar to \citet{bianchini16}). Separating the sub-population into 10 mass bins containing an equal number of stars, we measure the velocity dispersion $\sigma$ of all stars within the bin. Similar to our calculation of $\delta_\alpha$, the mean mass of all stars within the bin is used to set m. $\eta$ is then equal to the slope of the line of best fit to log($\sigma$) versus log($m$). 

Since $\eta$ is expected to vary with distance from the cluster center \citep{trenti13,bianchini16}, we have also explored the time evolution of $\eta$ and $\delta_\alpha$ including only stars within the half-mass radius. This constraint provides further insight into the evolution towards energy equipartition of the inner regions of a cluster. To compare with observational datasets, the calculations of both $\eta$ and $\delta_\alpha$ are repeated using each stars projected clustercentric radius ($R$) when radially binning the data and projected velocities to measure $\sigma$. And finally, to increase the number of stars used in each measurement of $\eta$ and $\delta_\alpha$ we group time steps in collections of three.

\subsection{Star Clusters with Different Tidal Filling Factors} \label{s_nobin}

\begin{figure}
\centering
\includegraphics[width=\columnwidth]{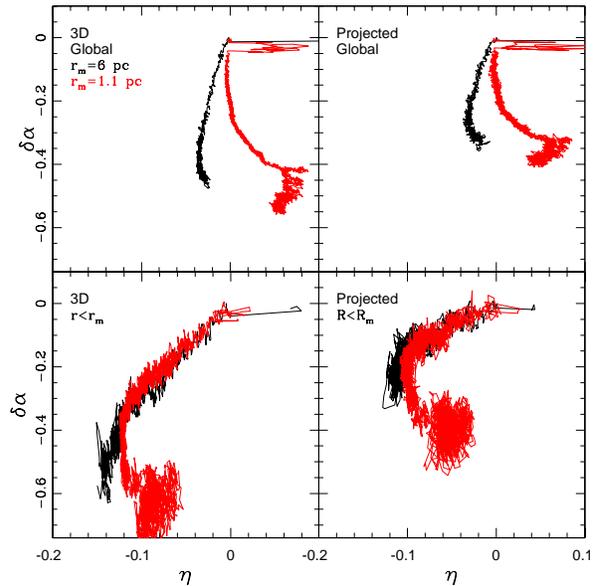}
\caption{Evolution of $\eta$ (for stars with $0.1 < m < 1.8 $) as a function of $\delta_\alpha$ (for stars with $0.1 < m < 0.5$) for stars in model clusters with circular orbits at 6 kpc in a Milky Way-like potential with $r_{m,i}=1.1$ pc (red) and $r_{m,i}=6$ pc (black). Upper Left Panel: $\eta$ and $\delta_\alpha$ are measured using the three dimensional position and velocity of all stars in the cluster. Lower Left Panel: $\eta$ and $\delta_\alpha$ are measured using the three dimensional position and velocity of stars with $r<r_m$. Upper Right Panel: $\eta$ and $\delta_\alpha$ are measured using the projected position and line of site velocity of all stars in the cluster. Lower Left Panel: $\eta$ and $\delta_\alpha$ are measured using the projected position and line of site velocity of stars with $R< R_m$.
}
\label{fig:tides}
\end{figure}

We begin our study by presenting the evolution of $\eta$ as a function of $\delta_\alpha$ for the $6\times10^4 M_\odot$ model clusters with $r_{m,i}=1.1$ pc and $r_{m,i}=6$ pc, K93 IMFs, and circular orbits at 6 kpc in a Milky Way-like potential (Figure \ref{fig:tides}). As previously mentioned, the $r_{m,i}=1.1$ pc is initially tidally under-filling while the $r_{m,i}=6$ pc cluster is initially tidally filling. Each parameter is presented in three dimensions and in projection. Additionally, $\eta$ and $\delta_\alpha$ are calculated using both the entire stellar population and only for stars with $r<r_m$ (or $R<$ Projected half-mass radius ($R_m$) when considering clusters in projection) in order to compare inner and global cluster properties. 

%The evolution of $\delta_\alpha$ and $\eta$ of both model clusters within a tidal field is nearly identical to the isolated case, with one minor difference. First, since these clusters are initially more massive it takes nearly twice as long for the $r_{m,i}=1.1$ pc cluster to reach core collapse. Therefore, it reaches a more negative value of $\delta_\alpha$ globally before $\eta$ begins to slow down and start increasing. However up until just before core collapse, the linear relationship between $\delta_\alpha$ and $\eta$ is the same as in the isolated case for stars with $r<r_m$,

Considering first the three-dimensional evolution of $\eta$ and $\delta_\alpha$ using all stars in the cluster (Figure \ref{fig:tides}: Upper Left Panel), we see that $\eta$ and $\delta_\alpha$ initially decrease, as expected, for both the the $r_{m,i}=1.1$ pc and $r_{m,i}=6$ pc clusters. Early fluctuations in $\eta$ are simply the result of remnants that have been given large velocity kicks which quickly escape the cluster. However it is not long before the evolution of both parameters in the two model clusters diverges, with $\eta$ actually increasing while $\delta_\alpha$ continues to decrease for the $r_{m,i}=1.1$ pc cluster. Eventually, once $\eta$ and $\delta_\alpha$ decrease a little further, $\eta$ starts increasing for the $r_{m,i}=6$ pc cluster as well. Once the $r_{m,i}=1.1$ pc cluster undergoes core collapse, the evolution of $\delta_\alpha$ slows dramatically while $\eta$ fluctuates about a mean value.

At first, an increase in $\eta$ seems counterintuitive as clusters are supposed to be evolving towards a state of energy equipartition ($\eta \sim -0.5$). However, this discrepancy can be understood by remembering that 1) $\sigma$ decreases with $r$, and 2) mean stellar mass starts to decrease with $r$ as the cluster undergoes mass segregation. Hence, while energy equipartition can be on-going locally in a cluster, the combined effects of mass segregation and tidal stripping can cause the global value of $\eta$ to increase with time. Since the under-filling cluster is able to expand more than the filling cluster, its velocity dispersion profile will be steeper and it will require less mass segregation (i.e a less negative $\delta_\alpha$) than the extended cluster before the global $\eta$ starts to increase. Similar effects are seen when $\eta$ and $\delta_\alpha$ are measured in projection (Figure \ref{fig:tides}: Upper Right Panel), albeit to a lesser degree since line of site velocity dispersion does not depend as strongly on r. 

To minimize the combined effects of local relaxation, mass segregation, and the radial dependence of $\sigma$ on $\eta$, we instead focus on stars within a narrower radial range. In the lower panels of Figure \ref{fig:tides} we consider the evolution of $\eta$ and $\delta_\alpha$ both in three dimensions (bottom left panel) and projection (bottom right panel) for stars with $r<r_m$ and $R<R_m$ respectively. In these lower panels we see that the co-evolution of $\eta$ and $\delta_\alpha$ is at first identical for the $r_{m,i}=1.1$ pc and $r_{m,i}=6$ pc clusters, with both parameters decreasing at similar linear rates. Both model clusters break from this linear relationship when $\delta_\alpha \sim -0.35$ and $\eta \sim -0.12$, as both parameters slow down due to cluster expansion with the local $t_{rh}$ increasing throughout the cluster. Therefore the fact that clusters reach a minimum value of $\delta_\alpha$ (as found by \citet{webb16}) is directly linked to a clusters ability to only reach a state of partial energy equipartiion \citep{trenti13, bianchini16}. For the more compact cluster with $r_{m,i}=1.1$ pc, $\eta$ eventually starts increasing while $\delta_\alpha$ continues to decrease. Both parameters eventually reach final constant values in each model.

% Quantify t split? It seems to be ~4 trh,2000 for each cluster

\begin{figure}
\centering
\includegraphics[width=\columnwidth]{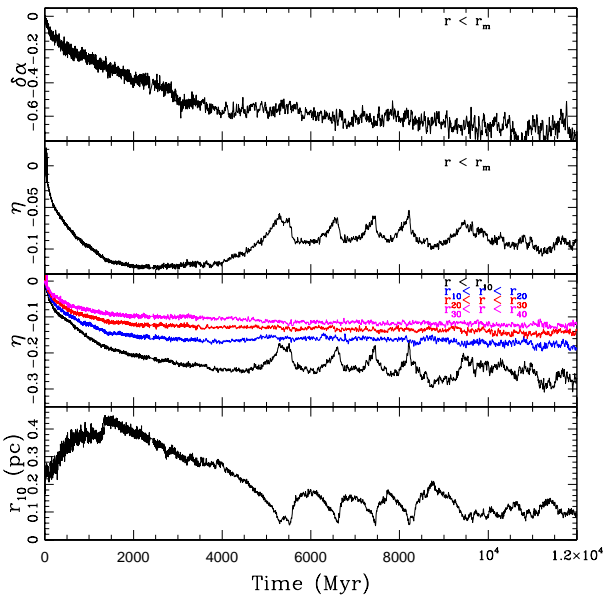}
\caption{Evolution of $\delta_\alpha$ for stars within $r_m$ and $0.1 < m < 0.5$ (Top Panel), $\eta$ for stars within $r_m$ and $0.1 < m < 1.8 $ (second panel from the top),  $\eta$ for stars within different Lagrange radii (second panel from the bottom) and the $10\%$ Lagrange radius (Bottom Panel) as a function of time for a model cluster with  $r_{m,i}=1.1$ pc and a circular orbit at 6 kpc in a Milky Way-like 
potential.}
\label{fig:tidesrh2}
\end{figure}

To understand why both model clusters break from the linear $\delta_\alpha$ - $\eta$ relationship, we consider the evolution of both parameters for all stars in the $r_{m,i}=1.1$ pc model with $r<r_m$ with time in Figure \ref{fig:tidesrh2}. For comparison purposes we also consider the evolution of $\eta$ within the $10\%$, $20\%$, $30\%$ and $40\%$ Lagrangian radii shells (as indicated in the legend) as well as the evolution of the $10\%$ Lagrange radius itself. The top and bottom panels of Figure \ref{fig:tidesrh2} illustrate that $\delta_\alpha$ within $r_m$ will decrease with time until the cluster approaches core collapse (at approximately 5.5 Gyr) and then remains more or less constant (in agreement with \citet{giersz96}). As the cluster continues to undergo repeated core collapse events ,$\delta_\alpha$ will resume decreasing towards its final minimum value (albeit at a much slower rate) as the cluster decreases in size while it slowly dissolves. $\eta$ within $r_m$ on the other hand stops decreasing much sooner, around 2.5 Gyr, despite the fact that $\eta$ is still decreasing locally within individual radial bins. This behaviour suggests that once a cluster reaches a high enough degree of mass segregation ($\delta_\alpha \sim -0.35$), the radial dependence of $\sigma$ and mean stellar mass will even start to affect measurements of $\eta$ over reduced radial ranges and $\eta$ within $r_m$ will stop decreasing and start increasing in the same way the global $\eta$ does. Therefore it is the combined effects of energy equipartition, mass segregation, and the radial dependence of $\sigma$ that cause the relationship between $\delta_\alpha$ and $\eta$ for stars within $r_m$ to break from linearity at later times. Below $\delta_\alpha \sim -0.35$, $\eta$ does not stop decreasing as quickly in the $r_{m,i}=6$ pc model because the radial dependence of $\sigma$ is not as steep as the $r_{m,i}=1.1$ pc model.

The increase in $\eta$ for stars with $r<r_m$ becomes more rapid as the cluster approaches core collapse. The increase in $\eta$ for all stars with $r<r_m$ is directly dependent on the behaviour of $\eta$ in the core of the cluster, as massive stars preferentially interact with newly formed binary stars and their velocity dispersion increases. Hence, as the cluster approaches core collapse $\eta$ is actually increasing locally within the core, and is not just a result of mass segregation and the dependence of $\sigma$ on r. After the initial core collapse, $\eta$ will fluctuate around a mean value as the cluster undergoes subsequent core collapse events, indicating that the mass segregation / energy equipartition process halts or has significantly slowed down in agreement with what is found in \citet{giersz96} \& Heggie (1996).    

\subsection{Star Clusters with Different IMFs}

%Salpeter235 is more evolved because it starts lower mass, and has a lower relaxation trh,2000 than the other clusters.

\begin{figure}
\centering
\includegraphics[width=\columnwidth]{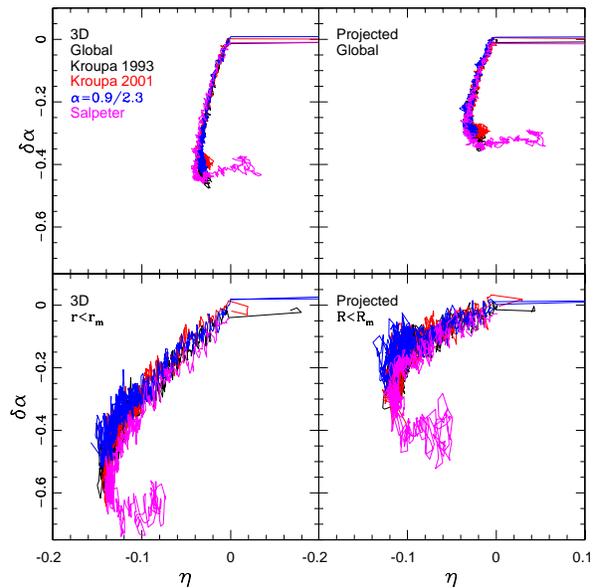}
\caption{Evolution of $\eta$ (for stars with $0.1 < m < 1.8 $) as a function of $\delta_\alpha$ (for stars with $0.1 < m < 0.5$) for stars in model clusters with circular orbits at 6 kpc in a Milky Way-like potential with a K93 IMF (black), K01 IMF (Red), broken power law IMF (blue) and Salpeter IMF (magenta). Upper Left Panel: $\eta$ and $\delta_\alpha$ are measured using the three dimensional position and velocity of all stars in the cluster. Lower Left Panel: $\eta$ and $\delta_\alpha$ are measured using the three dimensional position and velocity of stars with $r<r_m$. Upper Right Panel: $\eta$ and $\delta_\alpha$ are measured using the projected position and line of site velocity of all stars in the cluster. Lower Left Panel: $\eta$ and $\delta_\alpha$ are measured using the projected position and line of site velocity of stars with $R<R_m$.
}
\label{fig:IMF}
\end{figure}

To explore the effects of having a different IMF on the co-evolution of $\delta_\alpha$ and $\eta$, we make use of simulations with $r_{m,i} = 6$ pc as in the tidally filling clusters above but with four different IMFs from \citet{webb16}. We consider a K01 IMF, Salpeter IMF, and a broken power-law IMF with $\alpha$ equals -0.9 for $0.1 < m < 0.5 M_\odot$ and -2.3 for $0.5 \le m \le 50  M_\odot$. From Figure \ref{fig:IMF} we see that the co-evolution of $\delta_\alpha$ and $\eta$ is IMF-independent, as all four models have very similar evolutionary tracks in three dimensions and in projection. Therefore the relationship between $\delta_\alpha$ and $\eta$, including the linear relationship observed for stars with $r<r_m$ and $R < R_m$, is unaffected by whether or not the IMF is universal. A cluster's IMF, however, can influence where it is located on the $\delta_\alpha$ - $\eta$ evolutionary track at a given time if the cluster's structural evolution is also strongly affected. For example, the model with a Salpeter IMF will have a lower initial mass and expand less due to stellar evolution than the other clusters. These factors result in the Salpter IMF cluster having a shorter relaxation and reaching both the end of the linear evolutionary track (at $\delta_\alpha \sim -0.35$) and core collapse (causing $\eta$ to increase) within 12 Gyr.

\subsection{Star Clusters with Primordial Binaries, Different Orbits, and Different Black Hole Retention Fractions}

%Why are e0r18, e09rp6, and e0r104 less evolved? Is trh,2000 higher?
%Removed: While binaries have been shown to have a negligible affect on $\delta_\alpha$ \citep{webb16}, binaries could still possibly have an affect on the evolution of $\eta$. Similarily, while the evolution of $\delta_\alpha$ has been shown to respond to the structural changes that retaining black holes incurs on a cluster and is not directly affected by the black holes themselves \citep{webb16}, whether or not this statement applies to $\eta$ remains unknown. 

Finally, to study the evolution of $\delta_\alpha$ and $\eta$ over the widest possible range of initial conditions, we consider model clusters with different orbits, binary fractions, and black hole retention fractions taken from \citet{leigh13}, \citet{webb15}, and \citet{webb16}. Including model clusters with $r_{m,i}=1.1$ pc and $r_{m,i}=6$ pc and different orbital distances (6 kpc, 18 kpc, 104 kpc) allows for a large range of tidal filling factors to be explored. Including model clusters with $r_{m,i}=1.1$ pc and $r_{m,i}=6$ pc and having eccentric orbits (e=0.5 and 0.9) allows us to determine whether tidal heating and tidal shocks at perigalacticon (kept fixed at 6 kpc) affects the co-evolution of $\delta_\alpha$ and $\eta$.

\begin{figure}
\centering
\includegraphics[width=\columnwidth]{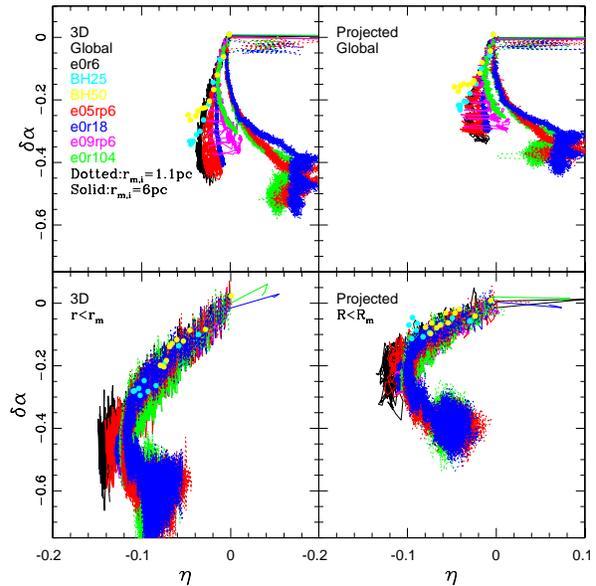}
\caption{Evolution of $\eta$ (for stars with $0.1 < m < 1.8 $) as a function of $\delta_\alpha$ (for stars with $0.1 < m < 0.5$) for stars in model clusters with circular orbits between 6 and 18 kpc, orbital eccentricities between 0 and 0.9, initial masses of $6 \times 10^4 M_\odot$, and initial half mass radii of 1.1 pc (dotted lines) and 6 pc (solid lines). Different model clusters are colour coded based on orbit, as indicated by the legend. Cyan and yellow data points correspond to model clusters with a circular orbit at 6 kpc and an initial half mass radius of 6 pc but with black hole retention fractions of $25\%$ and $50\%$ respectively. Upper Left Panel: $\eta$ and $\delta_\alpha$ are measured using the three dimensional position and velocity of all stars in the cluster. Lower Left Panel: $\eta$ and $\delta_\alpha$ are measured using the three dimensional position and velocity of stars with $r<r_m$. Upper Right Panel: $\eta$ and $\delta_\alpha$ are measured using the projected position and line of site velocity of all stars in the cluster. Lower Left Panel: $\eta$ and $\delta_\alpha$ are measured using the projected position and line of site velocity of stars with $R<R_m$. 
}
\label{fig:mw}
\end{figure}

We find that the presence of primordial binaries does not affect the evolution of either $\eta$ or $\delta_\alpha$, as model clusters in Figure \ref{fig:mw} with $r_{m,i}=1.1$ pc and $r_{m,i}=6$ pc and circular orbits at 6 kpc are identical to those in Figure \ref{fig:tides}. Conversely, the upper panels of Figure \ref{fig:mw} confirm our statements in Section \ref{s_nobin} that the co-evolution of $\delta_\alpha$ and $\eta$ is significantly affected by a cluster's initial relaxation time, tidal filling factor, tidal heating and tidal shocks. The dependence on initial relaxation time is related to how quickly a cluster undergoes core collapse, as mass segregation and energy transfer slow down and eventually stop during the post-core collapse phase. The dependence on the tidal filling factor can be seen by comparing clusters with similar initial relaxation times (i.e. initial sizes) in Figure \ref{fig:mw}. Since under-filling clusters expand more from their initial state than filling clusters, they will develop a steeper $\sigma$ profile and require less segregation before the global $\eta$ starts increasing. 

Model clusters with eccentric orbits are subject to tidal heating as they approach perigalacticon and a tidal shock at perigalactic, providing additional energy to stars orbiting in the outer regions of the cluster. Initially, before much segregation has taken place, tidal heating and shocks do not affect $\eta$ or $\delta_\alpha$ as the additional energy is dispersed among stars over the entire mass spectrum. However as $\delta_\alpha$ continues to decrease, the additional energy starts to primarily go to the low mass stars which come to dominate the outer region of the cluster. The energy injection leads to increasing the velocity dispersion of low mass stars, which in turn causes $\eta$ to briefly decrease when the cluster is near perigalacticon. However once these stars escape the cluster $\eta$ returns to its normal evolution. 

The strength of the external tidal field has a much lesser effect on stars in the core, such that $\eta$ in the inner regions of filling and under-filling clusters will evolve self similarly. This point is illustrated in the lower panels of Figure \ref{fig:mw} where the evolutionary tracks of all model clusters are nearly identical when only stars with either $r<r_m$ and $R<R_m$ are considered. For model clusters with identical initial relaxation times, the cluster's tidal filling factor  is able to influence its location on the $\delta_\alpha$ - $\eta$ evolutionary track as under-filling clusters are able to expand more than filling clusters such that their relaxation time become longer. Hence $r_{m,i}=6$ pc models e0r18, e09rp6, and e0r104 are just reaching the end of the linear portion of the $\delta_\alpha$ - $\eta$ evolutionary track after 12 Gyr while e0r6 and e05rp6 have evolved past it with $\delta_\alpha$ values less than -0.35.

To illustrate the effects of retaining black holes, we have also plotted in Figure \ref{fig:mw} model clusters from \citet{webb16} with $25\%$ and $50\%$ black hole retention fractions at evenly spaced time steps (cyan and yellow respectively) up until they have lost the same amount of mass as model e0r6. As discussed in \citet{webb16}, the retention of black holes slows the mass segregation processes such that $\delta_\alpha$ stops evolving at a higher (less negative) value in BH25 and BH50 than in e0r6 (also in agreement with \citet{trenti10, lutzgendorf13}). Figure \ref{fig:mw} illustrates that $\eta$ also behaves in a similar fashion. Hence the co-evolution of $\delta_\alpha$ and $\eta$ is unaffected by the presence of black holes in a star cluster up until the cluster approaches dissolution.

\section{Discussion and Summary}\label{s_discussion}

Using a suite of $N$-body simulations, we have found that the co-evolution of a cluster's global $\eta$ and global $\delta_\alpha$ can be strongly affected its initial relaxation time and tidal filling factor. $\delta_\alpha$ has previously been shown to decrease faster for clusters with shorter relaxation times, with a strong external tidal field having the capability of slowing $\delta_\alpha$'s rate of decrease by stripping preferentially low mass stars from the cluster once some segregation has occurred \citep{webb16}. And while the global $\eta$ does initially decrease as expected, the combined effects of local relaxation, mass segregation, and the radial dependence of $\sigma$ due to tidal stripping can all cause the global $\eta$ to actually start increasing. This apparent increase makes the global $\eta$ a poor indicator of the degree of energy equipartition in a cluster. 
 
The relationship between $\eta$ and $\delta_\alpha$ for stars with $r<r_m$ and $R<R_m$ on the other hand is linear and appears to be the same for all clusters independent of their initial conditions. The behaviour of both parameters is less affected by the strength of an external tidal field. Only strongly  segregated clusters, characterized by a steep radial dependence in $\sigma$ and mean stellar mass, and clusters that are undergoing or have undergone core collapse (such that mass segregation / energy equipartition has ended) will evolve away from this relation.

\begin{figure}
\centering
\includegraphics[width=\columnwidth]{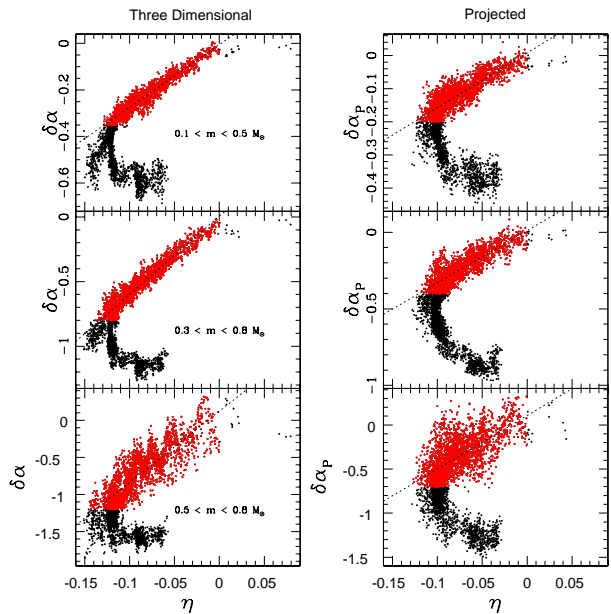}
\caption{Evolution of $\eta$ (for stars with $0.1 < m < 1.8 $) as a function of $\delta_\alpha$ (measured over different mass ranges) for stars within $r_m$ of model clusters with circular orbits between 6 and 104 kpc, orbital eccentricities between 0 and 0.9, initial masses of $6 \times 10^4 M_\odot$, and initial half mass radii of 1.1 pc and 6 pc. In the left column the three dimensional positions and velocities are used to measure $\eta$ and $\delta_\alpha$ while in the right column the projected values of these quantities are used. In the top row only stars with $0.1 < m < 0.5 M_\odot$ are used to calculate $\delta_\alpha$ while in the middle and bottom columns stars $0.3 < m < 0.8 M_\odot$ and $0.5 < m < 0.8 M_\odot$ are used respectively. Red points show the values used for the calculation of the linear best-fit line shown as a dotted line in each panel. The numerical values of the parameters of each best-fit line are reported in Table \ref{table:fitparam}.}
\label{fig:fit}
\end{figure}

To quantify the linear portion of the relationship between $\eta$ and $\delta_\alpha$ for stars within $r_m$, we plot in Figure \ref{fig:fit} the $\eta$ and $\delta_\alpha$ (measured using stars with $r<r_m$ (left panel) and $R<R_m$ (right panel)) for each model cluster in this study. We then fit a line of best fit to the data points which fall along the linear portion of each plot (highlighted in red) and measure its slope and y-intercept. As previously discussed, the linear portion of the track ends when $\delta_\alpha$ (when measured using stars between 0.1 and 0.5 $M_\odot$) reaches -0.35 (-0.2 in projection), which causes $\eta$ within $r_m$ to stop decreasing due to the dependence of both mean stellar mass and $\sigma$ on clustercentric distance. In order for our results to be applicable to a wide range of observational studies, the process is repeated for values of $\delta_\alpha$ measured using stars with $0.3 < m < 0.8 M_\odot$ (middle row) and $0.5 < m < 0.8 M_\odot$ (lower row). Over these mass ranges, $\delta_\alpha$ reaches -0.8 and -1.2 (-0.4 and -0.7 in projection) respectively by the time $\eta$ stops decreasing. We report in Table \ref{table:fitparam} the slope and intercept (along with the error on the values of these parameters) of the best-fit line found via linear regression of each of the panels shown in Figure \ref{fig:fit}, as well as the coefficient of determination $R^2$ for each fit and the minimum value of $\delta_\alpha$ for which the fits are applicable.

%Do we want to use sigma here even though its velocity dispersion?
%Do we want to use chi^2? Or reduced chi^2?

\begin{table*}
  \caption{Lines of Best Fit}
  \label{table:fitparam}
  \begin{center}
    \begin{tabular}{lcccc}
      \hline\hline
      {Mass Range} & {Slope} & {y-intercept} & $R^2$ & $\delta_{\alpha,min}$ \\
      \hline

$0.1 < m < 0.5 M_{\odot}$ &{}&{}&{}&{} \\
3D & 2.69 $\pm$ 0.02 & 0.005 $\pm$ 0.002 & 0.92 & -0.35 \\
2D & 1.65 $\pm$ 0.02 & 0.008 $\pm$ 0.002 & 0.73 & -0.2 \\

$0.3 < m < 0.8 M_{\odot}$&{}&{}&{}&{} \\
3D & 6.05 $\pm$ 0.03 & 0.005 $\pm$ 0.003 & 0.92 & -0.8 \\
2D & 3.41 $\pm$ 0.05 & 0.011 $\pm$ 0.004 & 0.72 & -0.4 \\
    
$0.5 < m < 0.8 M_{\odot}$&{}&{}&{}&{} \\
3D & 9.6 $\pm$ 0.1 & 0.12 $\pm$ 0.01 & 0.74 & -1.2 \\
2D & 6.1 $\pm$ 0.1 & 0.11 $\pm$ 0.01 & 0.46 & -0.7 \\

      \hline\hline
    \end{tabular}
  \end{center}
\end{table*}

Figure \ref{fig:fit} and Table \ref{table:fitparam} provide a direct link between energy equipartition and mass segregation (measured via radial variation in the stellar mass function) in star clusters. They also prove that the minimum value of $\delta_\alpha$ reached by the models of \citet{webb16} is the result of clusters only reaching a state of partial energy equipartition ($\eta < -0.5$) \citep{trenti13, bianchini16}. The linear relationship between $\eta$ and $\delta_\alpha$ within $r_m$ offers the possibility of inferring the kinematic properties of a star cluster from its photometric properties (and viceversa). The distribution of stellar masses in a cluster can be used to provide an indication of $\eta$ without the need for spectroscopy or proper motions.  Furthermore, if stellar masses and velocities are available within the same globular cluster, then clusters which do not fall on the linear relationship could potentially be identified as core-collapse clusters.

One factor that has not been explored by our suite of simulations is the effect of primordial mass segregation on the coevolution of $\eta$ and $\delta_\alpha$. The effects of primordial mass segregation on $\delta_\alpha$ were considered in \citet{webb16}, where it was determined that while primordial mass segregation alters the initial value of $\delta_\alpha$ at time zero, a primordially mass segregated cluster will reach a similar $\delta_\alpha$ than a non-primordially segregated cluster within a few relaxation times. What needs to be considered for the purposes of this study however is whether or not primordial mass segregation is linked to mass-dependent kinematical properties. Primordial mass segregation might be in part driven by early dynamics, it could be mainly due to the star formation process or a combination of star formation and early dynamics \citep{bonnell01,mcmillan07, moeckel10, maschberger10}; in any case the relation between spatial segregation and the kinematical properties is in general non-trivial and its investigation is well beyond the scope of this study. 

Our work provides a theoretical prediction for the relationship between $\eta$ and $\delta_\alpha$, and can be used today to estimate $\eta$ for clusters that have an accurately measured $\delta_\alpha$ via wide field studies. With upcoming proper motion measurements from the Hubble Space Telescope and GAIA, it is becoming possible to actually measure $\eta$. Hence the predicted relationship between $\eta$ and $\delta_\alpha$ will soon be able to be observationally tested. Observational verification would serve as a clear indication that the initial conditions and long-term evolution of star clusters are well understood and strengthen the applicability of the $\eta$ / $\delta_\alpha$ relation as a tool for studying globular clusters. A disagreement between theory and observations on the other hand would indicate additional studies are necessary to determine what types of initial conditions (including primordial mass segregation) or dynamical processes could alter the co-evolution of $\eta$ and $\delta_\alpha$ found here.

\section*{Acknowledgements}

This work was made possible in part by the facilities of the Shared Hierarchical Academic Research Computing Network (SHARCNET:www.sharcnet.ca) and Compute/Calcul Canada, in part by Lilly Endowment, Inc., through its support for the Indiana University Pervasive Technology Institute, and in part by the Indiana METACyt Initiative. The Indiana METACyt Initiative at IU is also supported in part by Lilly Endowment, Inc.

\bsp

\label{lastpage}

\end{document}